\documentclass[useAMS,usenatbib,usegraphicx]{mn2e}

\def\lyba{L\lowercase{y}${\balpha}$}
\def\lya{L\lowercase{y}${\alpha}$}

\def\microns{{$\mu$}m}

\def\z9{$z\sim9$}

\title[A deep, narrow $J$--band search for proto--galactic {\lyba}
 emission at redshifts {\z9}.]{A deep, narrow $J$--band search for
 proto--galactic {\lyba} emission at redshifts {\z9}\thanks{Based on
 observations collected at the European Southern Observatory, Chile
 (ESO Programme 69.A--0330).}.}

\author[J. P. Willis and
F. Courbin]{J. P. Willis$^{1}$\thanks{E-mail:jwillis@uvic.ca} and
F. Courbin$^{2}$\\ $^{1}$Department of Physics and Astronomy,
University of Victoria, Elliot Building, 3800 Finnerty Road, Victoria,
BC, V8P 1A1, Canada.\\ $^{2}$ Ecole Polytechnique F\'ed\'erale de
Lausanne (EPFL), Laboratoire d'astrophysique, Observatoire, CH-1290
Chavannes-des-Bois, Switzerland.\\}

\begin{document}

\date{Accepted ????. Received ????; in original form ????}

\pagerange{\pageref{firstpage}--\pageref{lastpage}} \pubyear{????}

\maketitle

\label{firstpage}

\begin{abstract}

We present a deep, narrow $J$--band search for proto--galactic {\lya}
emission at redshifts {\z9}. We combine an exceptionally deep image of
the Hubble Deep Field South, obtained using a narrow band filter
centred on the wavelength 1.187{\microns} using the VLT/ISAAC
facility, with existing deep, broad band images covering optical to
near infrared wavelengths. We search for \z9\ \lya--emitting galaxies
displaying a significant narrow band excess relative to the
$J_s$--band that are undetected at optical wavelengths. We detect no
sources consistent with this criterion to the 90\%\ point source flux
limit of the NB image, $F_{NB} = 3.28 \times 10^{-18} \rm \, ergs \,
s^{-1} \, cm^{-2}$. The survey selection function indicates that we
have sampled a volume of approximately $340 \, h^{-3} \, \rm Mpc^3$ to
a \lya\ emission luminosity of $10^{43} h^{-2} \rm \, ergs \, s^{-1}
$. We conclude by considering the potential implications for the
physics of the high--redshift universe.

\end{abstract}

\begin{keywords}
galaxies: high redshift
\end{keywords}

\section{Introduction}

\subsection{The high--redshift universe}

Studies of galaxies and QSOs at the highest known redshifts have
historically provided some of the strongest constraints upon theories
of galaxy formation and the physical state of the Inter--Galactic
Medium (IGM). Observations of the high--redshift universe have been
extended progressively during the past decade via a series of
dedicated studies: \citet{steidel96} demonstrated that optical colour
selection techniques, tuned to identify the spectral signature of
young star forming galaxies attenuated by intervening neutral
hydrogen, are highly effective at identifying large numbers (hundreds
per square degree) of galaxies at redshifts $2.5 < z <
4.5$. Spectroscopic observations of large samples (approximately 1000
to date) of so--called `Lyman--Break' galaxies has permitted
computation of the galaxy luminosity function \citep{steidel99},
composite stellar populations (\citealt{pettini01};
\citealt{shapley01}) and the amplitude of large scale structure
variations \citep{giavalisco98}. Employing similar colour selection
techniques, samples of distant galaxies have been extended to
redshifts $z=6$, with the advent of the Advanced Camera for Surveys
(ACS) facility deployed on the Hubble Space Telescope (HST;
\citealt{dickinson03}; \citealt{stanway04a}). Though such distant
surveys constrain the volume averaged star formation rate (SFR) for a
flux limited sample at early cosmic times \citep{gia03}, detailed
statistical conclusions await the compilation of larger samples.
Although the detection of a \lya--emitting galaxy at a redshift $z=10$
was recently reported by \citet{pello04} following a colour--selected
survey for redshift $z>7$ galaxies located behind well--studied
gravitational lens clusters, the exact nature of the source remains
contentious (e.g. \citealt{weather04}; \citealt{bremer04}).

Observational programmes employing narrow band (NB) imaging techniques
combine the advantage of wide area imaging surveys with enhanced
sensitivity to discrete spectral features. Though NB selection
techniques restrict such programmes to line--emitting galaxies at
specific redshifts, spectroscopic confirmation of sources
pre--selected to to display strong spectral features is relatively
straightforward compared to continuum selected samples. A number of NB
imaging programmes have successfully extended the study of
high--redshift objects from redshifts $z=3-4$ \citep{cowie98}, through
redshifts $z=4-6$ (\citealt{malhotra02}; \citealt{rhoads03};
\citealt{hu04}) to redshifts $z=6.5$ (\citealt{hu02a};
\citealt{hu02b}; \citealt{kodaira03}). Such studies have extended
current constraints upon the star formation history of the universe to
redshifts $z \sim 6$, albeit incorporating a bias toward emission line
sources (i.e. unobscured star formation).

Studies of high-redshift QSOs (of greater apparent brightness than
galaxies at the same luminosity distance yet presenting surface
densities lower by several orders of magnitude to a given flux limit)
have been revolutionised by the Sloan Digital Sky Survey (SDSS;
\citealt{fan03}).  Detailed study of the current sample of $\sim10$
QSOs identified at redshifts $z>5.7$ have provided compelling
evidence for the onset of neutrality of the hydrogen component of the
IGM \citep{becker01}, the existence of sustained metal production in
QSOs \citep{pentericci02} and the presence of massive black
holes in the universe approximately 0.9 Gyr after the epoch of
recombination \citep{fan01}.

\subsection{H\,{\sevensize I} absorption: the {\lya} forest and the IGM}
\label{sec_intro2}

Though evidence exists that star forming structures are present in the
universe at redshifts $z>6$, the possibility to observe such
structures is dependent critically upon the opacity of intervening
material.  The observed spectral energy distributions (SEDs) of high
redshift sources display Lyman series absorption arising from two
distinct phenomena: discrete structures and continuous, diffuse
absorption. Discrete H\,{\sevensize I} structures along the
line--of--sight to background sources each give rise to an individual
{\lya} absorption feature. As the absorption redshift increases, the
discrete `forest' of {\lya} absorption merges to create a blanketed
region of near contiuous absorption at wavelengths blueward of the
{\lya} transition in the background source. It is this absorption that
gives rise to the Lyman--Break discontinuity present in the SEDs of
all high--redshift sources.  Songaila and Cowie (2002) estimate the
amplitude of the Lyman Break discontinuity to be
\begin{equation}
\Delta m = 3.8 + 20.3 \log_{10} \left ( \frac{1+z}{7} \right ) ,
\end{equation}
from the average transmission of the {\lya} forest region in the
spectra of a sample of 15 QSOs located over the redshift interval
$4.42<z<5.75$.  However, extrapolation of this relation to redshifts
$z>6$ is complicated as the optical depth of the intervening IGM
potentially approaches and exceeds unity.  The high redshift IGM can
be approximated to a continuous spatial distribution of hydrogren
gas. A uniform IGM becomes opaque ($\tau \ge 1$) to wavelengths
$\lambda < 1216${\AA} at a neutral hydrogen fraction of $10^{-5}$
\citep{gunn65}. As the fraction of neutral gas increases, the
absorption saturates and develops a characteristic `damping wing' --
extending the absorption profile to wavelengths redward of {\lya} and
potentially attenuating {\lya} emission in the embedded source.
Spectroscopic observations of a subset of high--redshift QSOs
identified within the SDSS \citep{becker01} indicate the presence of
absorption troughs where the line--of--sight transparency is zero,
i.e. apparently indicating $\tau \ge 1$.  However, associating
extended regions of zero transmitted flux with the Gunn--Peterson (GP)
effect is complicated by the requirement to disentangle the effects of
line blanketing in the {\lya} forest from the excess absorption
arising from the onset of neutrality in the IGM itself (Songaila and
Cowie 2002).  Although the above observations may have detected
initial indications of the onset of neutrality in the IGM, it is
important to note that, due to density fluctuations in the IGM and
clustering of sources of ionising radiation, the apparent onset of
neutrality in the IGM may be a strong function of the line--of--sight.
Further observations are required to determine the extent of
variations in the properties of the IGM due to these sources of cosmic
variance \citep{white03}.  Furthermore, the observation of a {\lya}
emitting galaxy at redshift $z=6.56$ \citep{hu02a} is not inconsistent
with observation of a possible GP effect in lower redshift
QSOs. \citet{haiman02}, \citet{santos04} and \citet{barton04} describe
the ionising effect of a star forming galaxy embedded in a neutral IGM
and note that, depending upon the exact assumptions made regarding the
mass and star formation properties of the source and the physical
conditions present in the IGM, the galaxy will form a local
H\,{\sevensize I}{\sevensize I} region of sufficient size to permit
transmission of a partially attenuated {\lya} line and associated
continuum.

In contrast to observations of absorption troughs in redshift $z\sim6$
QSO spectra, observations of the Cosmic microwave Background (CMB)
have the potential to constrain the properties of the universe during
the reionisation epoch optical depth regime $\tau \gg 1$. Observation
of a positive correlation in the Wilkinson Microwave Anisotropy Probe
(WMAP) CMB polarisation--temperature map at large angular scales can
be interpreted as a scattering event with an optical depth $\tau =
0.17$ -- consistent with the reionisation of the universe at redshifts
$z<30$ \citep{kogut03}. Although the exact variation of the mean
neutral hydrogen fraction as a function of redshift is strongly model
dependent, the initial WMAP results support an early commencement of
the reionisation epoch.

The present paper describes a dedicated search for high--redshift star
forming galaxies, employing an extension of broad and narrow--band
selection techniques applied at optical wavelengths to the near
infrared (NIR) wavelength regime. In particular we focus upon the
application of a narrow $J$--band filter centred at
$\lambda=1.187${\micron} to detect the signature of {\lya} emitting
galaxies located about a redshift $z=8.8$ (termed \z9 in the following
text).  The present work builds upon earlier searches for galaxies at
extreme redshift employing a similar narrow $J$--band approach
\citep{parkes94}, in addition to narrow $K$--band studies potentially
sensitive to galaxies at even greater redshift \citep{bunker95}.
The following sections are organised as follows: in Section 2 we
describe the observational strategy, the choice of target field and we
describe the narrow $J$--band data acquisition and reduction. We
further describe the archival data employed and the creation of a
multi--colour catalogue for the field. In Section 3 we discuss source
detection, completeness and contamination within the multi--colour
catalogue. In Section 4 we describe candidate emission line objects
detected in the field and we present our conclusions is Section
5. Unless otherwise indicated, values of $\Omega_{\rm M,0} = 0.3$,
$\Omega_{\rm \Lambda, 0} = 0.7$ and ${\rm H_0} = 70$ kms$^{-1}$
Mpc$^{-1}$ are adopted for the present epoch cosmological parameters
describing the evolution of a model Friedmann--Robertson--Walker
universe. Where used, $h$ is defined as $h = H_0 / 100$ kms$^{-1}$
Mpc$^{-1}$.

\section{Observations}

\subsection{Observing strategy and field selection}

Following the discussion in Section \ref{sec_intro2} it is clear that
candidate redshift \z9 galaxies will display a well defined
photometric signature in a search employing optical to NIR broad band
photometry with a narrow band centred on {\lya} emission. The
combination of {\lya} forest and IGM absorption will result in
effectively zero flux transmission at rest wavelengths $\lambda <
1216${\AA}. However, uncertainty regarding the physical state of the
IGM and the potential for \z9 sources to create local H\,{\sevensize
I}{\sevensize I} regions ensures that a range of {\lya} emission
properties may result from a given star forming galaxy
population. Throughout this paper, we consider that a redshift \z9
{\lya}--emitting galaxy will display a significant narrow band excess
relative to the $J_s$--band, in addition to displaying a continuum
break consistent with almost complete attenuation of photons at rest
frame $\lambda < 1216${\AA}. In order to generate an effective survey
for such sources certain additional factors must be considered:
\begin{enumerate}

\item{NIR continuum imaging data must achieve a limiting depth of $\rm
AB \ge 25.5$ . The brightest emission line galaxies confirmed at
redshifts $z=5.7$ display AB magnitudes\footnote{Source magnitudes are
computed using the AB system \citep{oke74}, i.e. $\rm AB = -48.6 - 2.5
\log F_\nu$, where $\rm F_\nu$ is the spectral energy density within a
particular passband in units of $\rm ergs \, s^{-1} \, cm^{-2} \,
Hz^{-1}$.} $z^{\prime} \approx 24.5-25$ \citep{hu04}\footnote{Note
that we employ this comparison as, at redshifts $z=5.7$ and $z=8.8$,
redshifted H\,{\sevensize I} absorption lies at the blue edge of the
$z^{\prime}$ and $J_s$ bandpasses respectively.}. The additional
distance modulus between a redshift $z=5.7$ and $z=8.8$ results in an
additional dimming term of 1 magnitude.}

\item{Optical imaging data must reach a limiting depth typically 1.5
magnitudes fainter than NIR data. Early--type galaxies located at
redshifts $z \sim 2$ can generate a spectral discontinuity between
optical and NIR continuum bands of amplitude $D \sim 1.5$ mag.
\citep{stanway04b}. Failure to identify the continuum break directly
could lead to the mis--identification of redshifted [OII]3727 emission
in such sources as candidate \z9 {\lya} emission.}

\end{enumerate}

Following these considerations the Hubble Deep Field South (HDFS;
\citealt{williams00}) Wide field Planetary Camera 2 (WFPC2) apex
pointing ($\alpha=22^h32^m55\fs464, \,
\delta=-60\degr33\arcmin05\farcs01$, J2000) was selected as the target
field in order to exploit the high quality of optical to NIR image
data available for the field. In particular, the combination of HDFS
WFPC2 and Very Large Telescope (VLT) Infrared Spectrometer And Array
Camera (ISAAC; \citealt{moorwood97}) observations of the field provide
images to typical depths AB=28 and AB=26 in optical and NIR bandpasses
respectively \citep{labbe03}.

\subsection{Narrow--band near infrared observations}

\begin{figure}
\includegraphics[width=84mm]{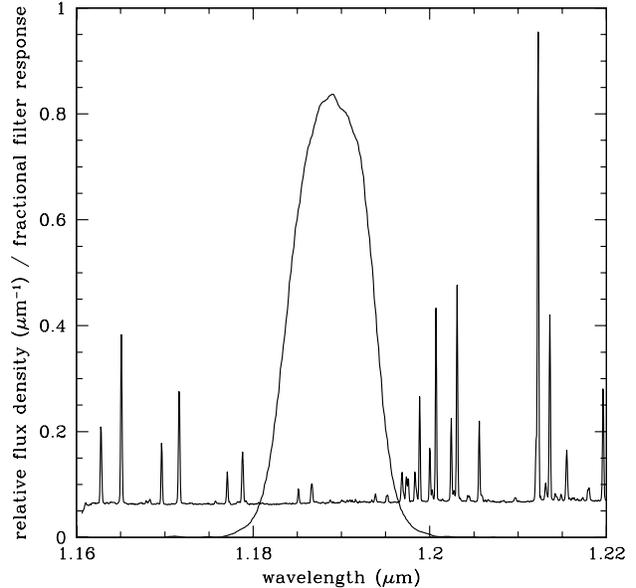}
\caption{The spectral response of the ISAAC NB119 filter superposed
upon the $J$--band emission spectrum of the night sky. The night sky
emission spectrum is taken from \citet{rousselot00}.}
\label{fig_nbplussky}
\end{figure}

Narrow $J$--band observations of the HDFS WFPC2 pointing were obtained
during ESO Period 69 (May 19th to September 17th 2002) employing the
ISAAC facility mounted on the Nasmyth--B focus of the 8.2 meter VLT
Antu telescope. The ISAAC short wavelength camera is equipped with a
Rockwell Hawaii $1024 \times 1024$ HgCdTe array. The pixel scale is
0\farcs1484 pix$^{-1}$ and the field size is $2\farcm5 \times
2\farcm5$. The read noise of the detector is 16.7 electrons and the
gain is 4.5 electrons ADU$^{-1}$. Observations were performed
employing the NB119 filter (Figure \ref{fig_nbplussky}). The filter
has an effective width of 89.5{\AA} and is centred at a wavelength
1.187{\microns} (corresponding to the location of {\lya} emission at a
redshift $z=8.76$). The NB119 filter is particularly effective for the
detection of faint emission as it is located in a region of low sky
noise, largely isolated from the `forest' of bright OH emission that
dominates the sky background in the $J$--band.

The science programme was allocated 32 hours of service mode
observations, to be obtained in seeing conditions of 0\farcs6 or
better. Observations were scheduled as 32 one hour observing blocks
(OBs), each consisting of ten, 300 second exposures dithered spatially
according to a Poissonian random offset located within a box size of
20{\arcsec} in right ascension and declination. The choice of pointing
centre and dither sequence were made to produce a final NB image of
the HDFS that was well matched to the available deep, broad NIR band
observations obtained as part of the Faint Infra--Red Extragalactic
Survey (FIRES) project \citep{labbe03}. The final data set consists of
42 hours of narrow band observations. This represents all usable
data, including observations obtained outwith the specified observing
constrains (sky transparency, moon phase and atmospheric seeing). The
data consists of 420 individual science images corresponding to a
total exposure time of 126,000 seconds.

\subsubsection{Data reduction}

Narrow band imaging data were a) corrected for varying pixel response
using twilight sky exposures, b) sky-subtracted having masked array
regions containing objects detected above a specified ADU level, c)
corrected for both high-- and low--frequency spatial artefacts, d)
shifted to a common pixel scale and coadded using a suitable pixel
rejection algorithm and bad pixel mask. The applied data reduction
techniques are broadly similar to those described in \citet{labbe03}
for deep $J_sHK_s$ observations of the HDFS and are summarised below.

Individual narrow band exposures were dark subtracted using the master
dark frame obtained during each observing night. The flat field
response and bad pixel masks were constructed using the {\tt flat}
routine in ECLIPSE\footnote{ECLIPSE is an image processing package
written by N. Devillard, and is available at
ftp://ftp.hq.eso.org/pub/eclipse/}. Twilight flat field exposures were
obtained employing a fixed exposure time and display uniformly
decreasing/increasing count levels throughout a given
sequence. Exposures displaying count levels less than 5000 ADUs were
excluded as computation of pixel responses based upon these frames
display a characteristic discontinuity at the location of the array
read junction indicating the presence of a residual dark
signature. The residual dark signature represents an additive effect
that can be minimised by constructing the flat field frame employing
twilight exposures displaying ADU levels greater than 5000.  The pixel
response is defined as the gradient of a linear regression fit to the
set of values returned by a given pixel over the twilight sequence
(the gain) normalised by the median gain of the array. Pixels
displaying gain values outwith the interval [0.5,2] were defined as
deviant pixels and added to the bad pixel mask.  Flat field frames
constructed from individual twilight sequences obtained over ESO
Period 69 were compared. The pixel-to-pixel root mean square (rms)
deviation between flat field frames is at the level of $0.2-0.4${\%}
per pixel. Large scale gradients in individual frames do not exceed
2{\%}. A master flat field frame was then constructed from all
suitable twilight exposures and employed to correct the pixel response
in the science exposures.

\begin{figure}
\includegraphics[width=84mm]{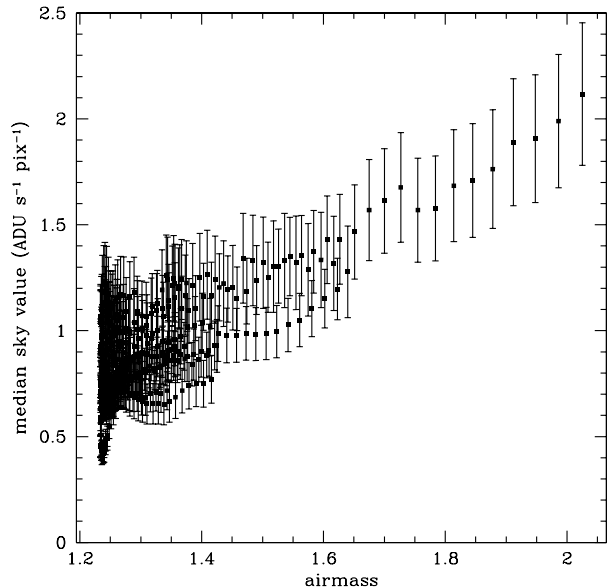}
\caption{Median sky value determined for 420 science frames as a
function of airmass computed at the mid--point of each frame. The
error bars represent the root mean square deviation of the sky pixel
distribution in each frame.}
\label{fig_sky_vs_airmass}
\end{figure}

The NB119 filter samples an exceptionally low sky background and
relatively long individual exposure times are required to ensure that
resulting images are limited by Poissonian sky noise and not by
detector read noise.  The median sky value in a typical 300 second
exposure obtained with the NB119 filter is 300 ADU pixel$^{-1}$ where
the exact value varies most significantly as a function of airmass
(Figure \ref{fig_sky_vs_airmass}).
Large variations in total sky emission or the
presence of gradients are not noted within a typical one hour
observing sequence (10 science frames). Accurate sky subtraction can
therefore be achieved employing standard techniques employing a
measure of the sky in a given frame drawn from a statistical
combination of science frames obtained adjacent in time. The sky
contribution in each pixel within a given science frame is computed
from the six adjacent frames in time. The sky contribution per pixel
is estimated via scaling the six frames to a common median level and
computing the average pixel value after rejection of the highest and
lowest values.

Pixel values in the sky image created for each frame are biased to
higher ADU levels by the contribution of astronomical
sources. Therefore, the location of bright objects within individual
images must be identified with an object mask and removed from second
computation of the sky level. Relative integer pixel shifts generated
by the dithered observing sequence are computed using the relative
location of bright stars in the sky subtracted images. The 420
individual images are registered and co--added (averaged) having
rejected the 10 highest and lowest pixels to produce an intermediate
image. All astronomical sources displaying a peak ADU level greater
than a $5~\sigma$ threshold are identified within the co--added image
and a set of masklets are created corresponding to the location of all
objects in the dithered images.  The above sky subtraction procedure
is then repeated excluding pixels identified by the object masklet
associated with a particular frame.

Sky subtracted images contain low level residuals (of the order of a
few ADUs) at both high and low spatial frequencies. High frequency
residuals arise from a residual dark signature (also termed pattern
noise), the amplitude of which is related to gradients in the dark
signal, i.e. it is largest at low row numbers (1,2,...) and those
following the read junction (513, 514, ...). This signature was
removed by computing the median value of all pixels contributing to a
given row, having first masked both bad pixel and object
locations. Analysis of corrected images indicated a residual signature
associated with the columns of each image. Therefore, the residual
dark signature in each column was corrected for via application of the
same procedures. Low frequency residuals largely arise from imperfect
sky subtraction and are removed by fitting a slowly varying function
(2nd order cubic spline) along rows and columns incorporating bad
pixel and object masking.

At this stage, background corrected, sky subtracted images contain
cosmic ray events and additional bad pixels not identified within the
flat field computation. Cosmic rays are identified employing the
PHIIRS\footnote{Pat Hall's Infrared Imaging Reduction Software.}
routine {\tt crzap}: each image is median smoothed over a given
spatial scale and the result is subtracted from the original
image. Application of a sigma threshold based upon the sky noise in
each frame identifies the signature of cosmic ray events. Events thus
identified are added to the bad pixel mask associated with that frame.
Unidentified bad pixels remaining in each frame are identified via
analysis of the rms--normalised pixel distribution in each frame with
bad pixel and object masking applied. Pixels displaying absolute
rms--normalised values greater than six are added to the bad pixel
mask associated with the particular frame. This procedure typically
identifies $40-50$ additional bad pixels per ten frame observing
sequence. Images analysed in this manner were then registered
employing integer pixel shifts and co--added employing both a bad
pixel mask and a pixel rejection criteria that rejected the 10 highest
and lowest values associated with a particular pixel. A pixel
rejection criteria was still found to be necessary in order to exclude
deviant pixels associated with objects (11{\%} of the image area) from
the co--added image.

The quality of the coadded image was improved employing a weighting
scheme to optimize the image combination given variations in seeing,
sky transparency and background noise within individual
images. Individual shifted images, $G_i$, were combined to produce a
final image, $F$, employing a weighting scheme such that $F = \sum w_i
G_i$, where $w_i = z_i / ( v_i s_i^2 )$. The applied weighting scheme
respectively corrects for relative variations in sky transparency via
$z_i$, derived from the mean 5\arcsec\ diameter aperture flux of four
bright reference stars within the HDFS field, variations in background
noise via, $v_i$, the measured variance per pixel and variations in
atmospheric seeing, $s_i$, measured from the profiles of the same four
bright reference stars. The distribution of image weights are
normalised such that $\sum w_i = 1$. The Full--Width at Half--Maximum
(FWHM) of the Point Spread Function (PSF) in the reduced image is
0\farcs47. The dimensions of the final coadded image are 2\farcm9
$\times$ 2\farcm9. A science image was created from this frame based
upon pixels displaying weighted exposure times greater than a fraction
0.95 of the 126,000 on--sky exposure time. The dimensions of this
science image are 2\farcm2 $\times$ 2\farcm2.

\subsubsection{Flux calibrating the NB119 image}
\label{sec_fluxcal}

\begin{figure}
\includegraphics[width=84mm]{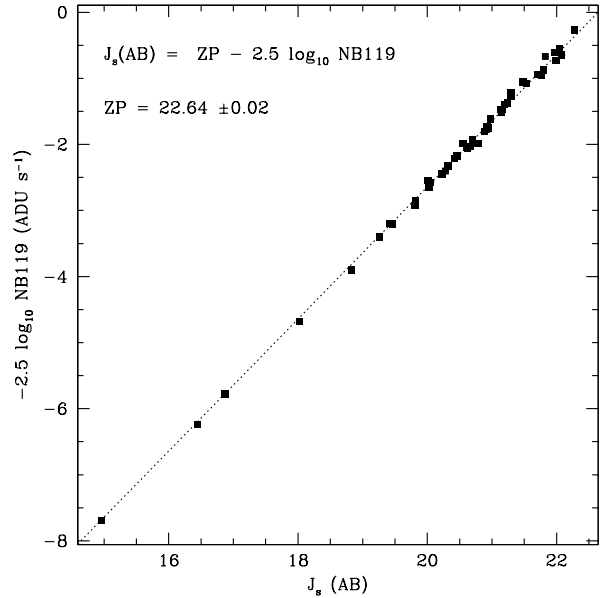}
\caption{Computation of the zero--point of the NB119 image. The dotted
line indicates a linear relationship of the form $y = x + \rm
ZP$. Individual error bars are displayed but possess a mean amplitude
0.006 magnitudes.}
\label{fig_fluxcal}
\end{figure}

The photometric zero point for the NB119 image was computed employing
47 bright, isolated sources common to the NB119 and FIRES $J_s$--band
images. Assuming that the SEDs of these calibration sources display no
strong discontinuities, the $J_s$--band flux density provides an
accurate estimate of the flux density with the NB119 filter,
i.e. $J_s-NB = 0$. Source detection was performed on the NB119 image
and apertures were transformed to the FIRES $J_s$ image to compute
$J_s$ magnitude values. The transformation was computed using the {\tt
IRAF}\footnote{IRAF is distributed by the National Optical Astronomy
Observatories, which are operated by the Association of Universities
for Research in Astronomy, Inc., under cooperative agreement with the
National Science Foundation.} routines {\tt geomap} and {\tt
geoxytran}. Twenty bright, nucleated galaxies were employed as
reference sources and the root mean square deviation of the reference
sources about the computed transformation was 0.2 pixels in the $x$--
and $y$--pixel directions. Source brightness measures were computed
within 5{\arcsec} diameter circular apertures and the distribution of
$J_s$ versus NB119 instrumental magnitude was found to be linear with
a gradient of unity. The relationship is displayed in Figure
\ref{fig_fluxcal} and indicates that 1 ADU s$^{-1}$ received in the
NB119 image corresponds to an AB magnitude of $22.64\pm 0.02$.

\subsection{Archival optical and near infrared observations}

Optical to NIR broad band photometry of the HDFS is taken from the
catalogue of \citet{labbe03}. The authors employ as input material for
the multi--colour catalogue seven images, corresponding to the
$U_{300}B_{450}V_{606}I_{814}J_sHK_s$ filters, that have each been
transformed and resampled to the astrometric system defined by the
F814W HST image and have been convolved with a kernel that reproduces
the seeing of the FIRES $H$--band image as determined via the median
FWHM of bright stars ($\rm FWHM = 0 \farcs 48$.). Photometric measures
are computed employing 0\farcs7 diameter apertures and are zero
pointed on the AB magnitude system. As noted in Section
\ref{sec_fluxcal}, the NB119 image is not transformed to the
astrometric system defined by the FIRES catalogue images as this would
require a resampling of the data and the unnecessary introduction of
correlated noise between adjacent pixels within our detection
image. Instead, the set of 0\farcs7 diameter detection apertures
generated by the NB119 image (see Section \ref{sec_complete} below)
are transformed to the FIRES astrometric system employing the
transformation described in Section \ref{sec_fluxcal}. Finally, no
additional kernel is applied to correct for the slight difference in
the measured seeing between the NB119 and FIRES $H$--band image as the
measured difference ($\rm \Delta FWHM = 0\farcs01$) is smaller than
the 0\farcs04 variation in seeing over the $H$--band image reported by
\citet{labbe03}.

\section{Source detection and completeness}
\label{sec_complete}

Source detection and analysis was performed on the NB119 image using
the {\tt SExtractor} software package \citep{bertin96}. Circular
apertures of diameter 0\farcs7 were employed to compute source
fluxes. Source detection criteria were tuned to select objects
displaying at least 12 pixels above a threshold of 0.4$\sigma_{sky}$.
Prior to source detection, the image was convolved with a `mexican
hat' filter of FWHM 0\farcs55.  As a band-pass filter, the mexican
hat emphasizes sources of extent comparable to its width, i.e. close to
the PSF extent in this case.  This was deemed an acceptable choice
given the assumption that target \z9 sources will appear unresolved in
both the NB119 and $J_s$ images.

The flux completeness limit of the NB119 image was estimated by
introducing and recovering artificial sources within the field. The
same analysis was performed upon the FIRES $J_s$--band image for
comparison.  Redshift \z9 sources are assumed to be unresolved in both
images.  The simulation proceeded by sub--dividing each image into
28$\times$28 cells of 5{\arcsec} side length.  Each of the 784 cells
was further divided into 256 regularly spaced points and an artificial
source was constructed at each point in the grid.  All simulated
sources in a particular image are introduced with the same source
profile and flux.  The image containing the artificial sources is then
added to the original image.  Simulated sources possess low brightness
values ($\rm AB > 24$) and their contribution to the total photon
noise per pixel is negligible.  Source extraction is then performed on
the image containing simulated sources. To avoid image blending among
the simulated sources, the source simulation and image addition plus
recovery operation is performed for each of the 256 locations within
the grid of 5{\arcsec} cells in the simulated frame. For each
5\arcsec\ cell the detection probability is the ratio of recovered to
introduced sources.

\begin{figure}
\includegraphics[width=84mm]{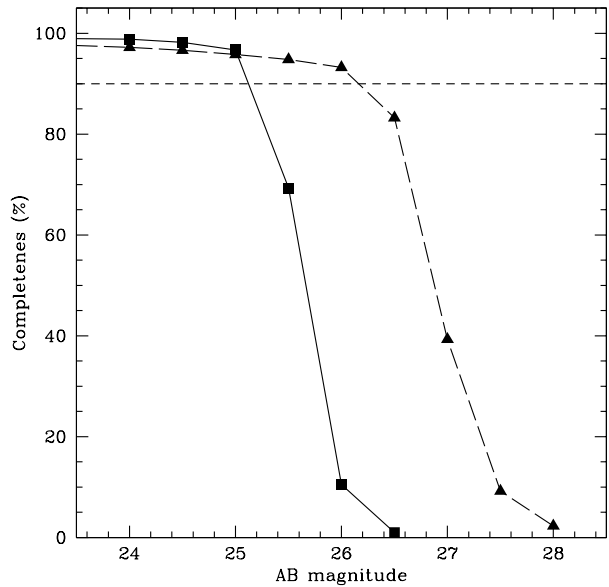}
\caption{Mean detection probability of simulated point sources as a
function of AB magnitude within the NB119 (square symbols) and $J_s$
images (triangular symbols). The horizontal dashed line indicates the
90\%\ completeness threshold.}
\label{fig_complete}
\end{figure}

Variations in the detection probability within a given map are
dominated by the local effect of bright objects within the image on
the likelihood to detect and extract simulated sources.  The
simulation of 256 source positions within each 5\arcsec\ cell
generates a grid of source centroids separated by 0\farcs3.  The
presence of a bright object within a particular 5\arcsec\ cell will
effectively mask all sources simulated in that cell and reduce the
corresponding detection probability. The effect is highly localised
and results in the apparently noisy aspect in the detection
probability maps.  These detection probability variations are of
considerable importance to any measure of the clustering of faint
sources in the field but do not compromise the estimation of the mean
completeness within a given source flux interval.  The mean value of
the detection probability at a given flux level, computed over the
entire field, corresponds to the completeness as a function of source
flux and is displayed for the NB119 and $J_s$ images in Figure
\ref{fig_complete}. Adopting the 90\%\ point source recovery threshold
as the limiting magnitude in each band generates magnitude limits of
$NB \le 25.2$ and $J_s \le 26.2$ respectively.  The rms amplitude of
sky counts measured in the NB image within the applied 0\farcs7
aperture corresponds to a magnitude $\rm AB=28.05$.  This figure is
approximately 20\%\ higher than the rms variation anticipated from the
extrapolation of background counts in raw data frames.  The integrated
signal--to--noise ratio (SNR) of an NB detection at $\rm AB=25.2$ is
therefore 13.8. The corresponding SNR value of a $J_s$--band detection
at $\rm AB=26.2$ is approximately 10.  The detectability of such
apparently significant sources is primarily affected by the additional
photon noise contributed by bright sources within the image area. We
note that the computed 90\%\ point source completeness limit of
$J_s=26.2$ agrees well with the corresponding value of $J_s=26.3$
derived by \citet{labbe03} given the slight differences between the
applied source simulation method and {\tt SExtractor} source detection
parameters in each case. In common with \citet{labbe03} we present
0\farcs7 aperture photometry throughout this paper. The correction
required to convert photometric measures computed in 0\farcs7
apertures to 5\arcsec\ apertures (which we assume to be `total'
measures) was determined to be 0.7 magnitudes via analysis of bright
stars in the NB image.

In addition to determining the completeness properties of the NB119
image, the false detection rate was also estimated. A set of 100
artificial images, containing no objects, and displaying random
background noise properties consistent with the final NB119 image, was
constructed. Source extraction was performed for each artificial image
and the mean false detection rate was computed.  With the combination
of {\tt SExtractor} parameters described above, a false detection rate
of 0.7 events per NB119 `noise' image resulted.

\section{Results}
\label{sec_results}

\begin{figure}
\includegraphics[width=84mm]{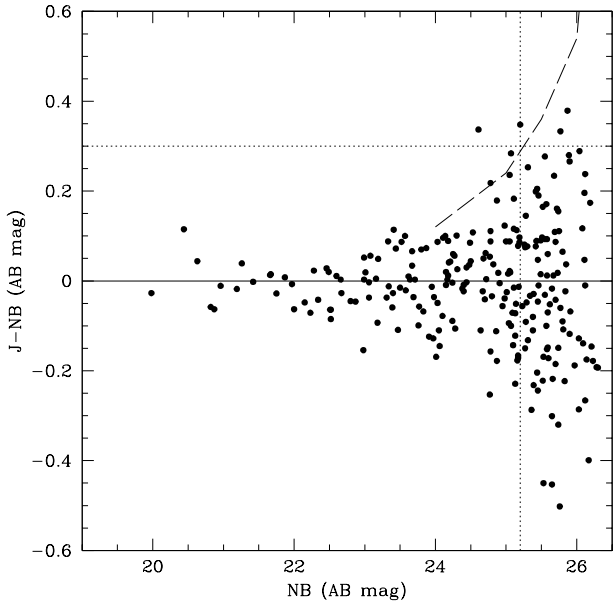}
\caption{Distribution of NB excess $J_s-NB$ versus $NB$ magnitude for
all objects in the NIR field area of the HDFS (solid points). The
solid horizontal line indicates $J_s-NB=0$. The vertical dotted line
indicates the selection criterion $NB\le25.2$. The horizontal dotted
line indicates the selection criterion $J_s-NB\ge0.3$ and corresponds
to a \lya\ EW of 27\AA\ (Appendix \ref{app}). The dashed curve
indicates the predicted $J_s-NB$ uncertainty ($3~\sigma$) as a
function of $NB$ magnitude returned by the completeness analysis
(Section \ref{sec_complete}).}
\label{fig_cmdresult}
\end{figure}

\begin{figure*}
\includegraphics[width=164mm]{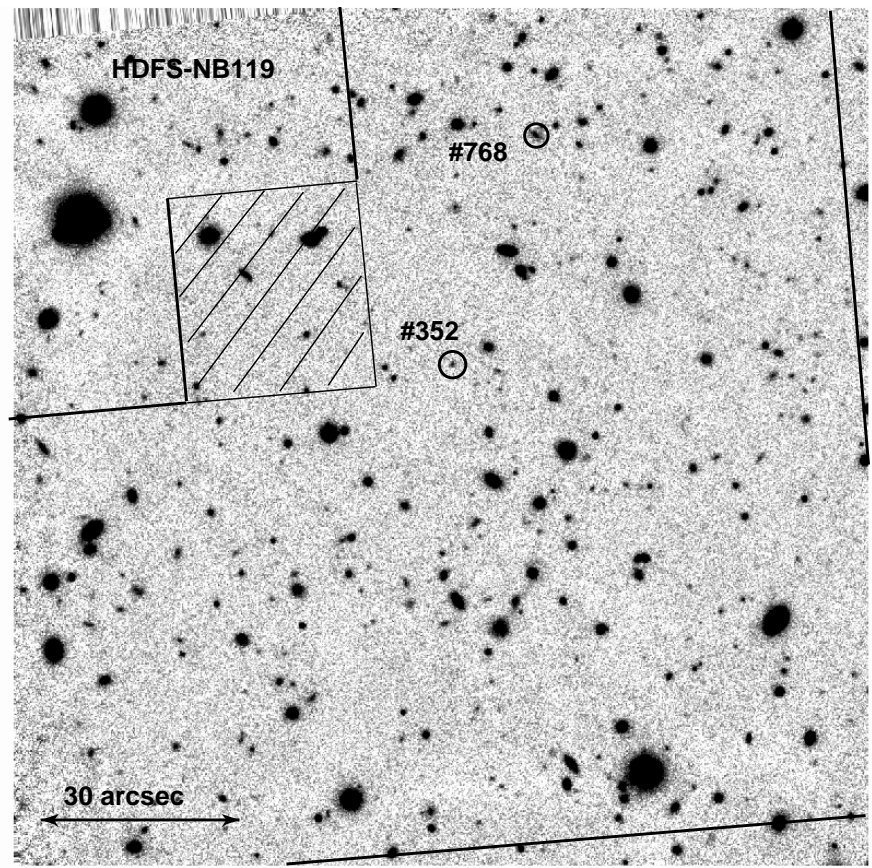}
\caption{Greyscale NB119 image of the HDFS. The image orientation is
North up and East left. The field geometry of the NIR broad band data
set is identical to the NB image. The field geometry of the HST WFPC2
field is indicated. Note that the Planetary Camera (shaded region) data
does not contribute to the final catalogue. The total field area
contributing to the NB excess catalogue is 4 arcmin$^2$. The location
of the two NB excess sources identified from the catalogue are
indicated.}
\label{fig_fieldhst}
\end{figure*}

Figure \ref{fig_cmdresult} displays NIR narrow band excess, expressed
as $J_s-NB$, versus $NB$ magnitude for all sources extracted from the
total area covered by the optical and NIR data for the HDFS field
(Figure \ref{fig_fieldhst}). Sources displaying $J_s-NB \ge 0.3$ and
$NB \le 25.2$ are flagged as potential \z9 emitting galaxies.  This NB
excess figure corresponds to a rest frame \lya\ equivalent width (EW)
of approximately 3\AA\ (Appendix \ref{app}) and represents a
$3~\sigma$ deviation from the flat spectrum hypothesis at the NB
magnitude threshold ($NB<25.2$) -- itself the 90\% point source
completeness limit (Figure \ref{fig_complete}).

\begin{figure}
\includegraphics[width=84mm]{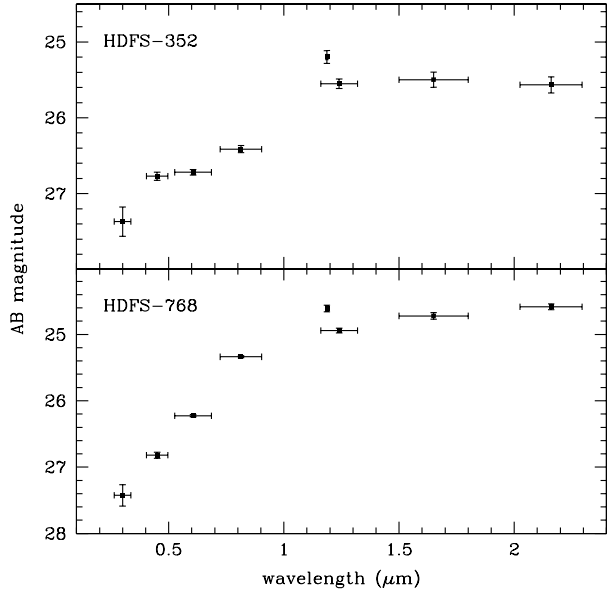}
\caption{Low resolution spectral energy distributions formed by the
multi--band photometry of the two NB excess objects identified within
the HDFS. The clear detection of each source within optical passbands
precludes each source as a candidate \z9\ source.}
\label{fig_phot_spec_2cand}
\end{figure}

Two sources are identified by the above selection criteria, HDFS--352
and HDFS--768, where the identification numbers refer to the HDFS
photometric catalogue of \citet{labbe03}. The photometric spectrum
formed by the $U_{300}B_{450}V_{606}I_{814}J_sHK_s$ plus NB photometry
of each object is displayed in Figure \ref{fig_phot_spec_2cand}. In
each case the addition of deep optical photometry is sufficient to
exclude each NB excess detection as a potential \z9\ source: each NB
excess object is detected in all optical bands and the photometric
redshift of each source \citep{rudnick01} -- $z_{phot} = 1.54 \pm
0.06$ (HDFS--352) and $z_{phot} = 0.76_{-0.20}^{+0.04}$ (HDFS--768) --
is consistent with a NB excess arising from redshifted H$\beta \,
4861$ and H$\alpha \, 6563$ emission respectively.
Although no candidate \z9\ sources are detected within the HDFS survey
area, the confirmation of two sources whose photometric redshifts are
consistent with the observed NB excess arising from narrow line
emission lends support to the assertion that the adopted selection
criteria identify faint, narrow emission line galaxies.

The \lya\ emission line selection function generated by the NB excess
search technique is described in Appendix \ref{app}. Although the
applied $NB\le25.2$ magnitude threshold corresponds to a total flux
integrated across the NB filter of $F_{NB} = 3.28 \times 10^{-18} \,
\rm ergs \, s^{-1} \, cm^{-2}$, computation of the flux transmitted
from a \lya--emitting galaxy located at a redshift $z$ and displaying
a rest frame velocity width $\sigma_v$, requires a more detailed
investigation of the applied selection criteria, the spectral response
of the NB119 filter and the assumed \lya\ source properties. Figure
\ref{fig_lv_samp} displays the co--moving volume sampled as a function
of \lya\ emission luminosity for three values of the rest frame
velocity width of the \lya--emitting source.  For the case where the
rest frame velocity width of putative \z9\ sources is $\sigma_v = 50
\rm \, kms^{-1}$, the NB excess survey area samples a co--moving
volume of $340 \, h^{-3} \rm \, Mpc^3$ to a \lya\ emission luminosity
of ${\rm L_{Ly\alpha}} \ge 10^{43} \, h^{-2} \rm \, ergs \,
s^{-1}$. 

\section{Conclusions}
\label{sec_conc}

No candidate \lya\ emitting sources at redshifts \z9\ have been
identified within the HDFS survey area by the selection criteria
described in Section \ref{sec_results}.  Expressing the observational
selection criteria in terms of the volume of the universe sampled to a
given \lya\ luminosity permits an investigation of this null result in
terms of the instrinic space density and luminosity of putative \lya\
emitting sources at \z9.  It must be noted however, that we currently
only consider the case where the \lya\ emission from \z9 sources is
unobscured, i.e. the escape fraction of \lya\ photons is unity.  This
scenario corresponds to the case where the \lya\ absorption opacity
due to the intervening IGM, and to the \lya\ emitting source itself,
is zero. Decreasing the relative fraction of \lya\ photons that escape
from \z9 sources will increase the corresponding upper limit on the
{\it unobscured} volume averaged luminosity density derived from the
NB excess survey. However, the degeneracy that exists between the
absorption properties of the IGM and the unobscured versus observed
volume averaged \lya\ luminosity density at \z9 complicates a more
complete investigation of the null result generated by the current NB
excess survey. The range of star formation conditions required to
maintain an ionised IGM at a redshift \z9\ remains a subject of active
discussion (c.f. \citealt{haiman02}; \citealt{santos04}) and is not
considered further in this paper.

A more empirical approach is to consider known \lya--emitting sources
and to transpose their properties to redshift \z9.  \citealt{hu04}
(hereafter H04) present a search for \lya\ emitting galaxies at $z
\sim 5.7$ employing a narrow band filter centred on the wavelength
8150\AA. Though the H04 study does not present the highest redshift
\lya\ sources currently known (c.f. \citealt{kodaira03}), their study
is based upon near complete spectroscopic follow--up observations --
19 out of 23 observed sources (from a total sample of 26 candidates)
were confirmed as \lya\ at $z=5.7$ -- and the overall design of their
study is very close to the NIR selected survey presented in this
paper.  H04 report a surface density of \lya\ emitting galaxies of
0.03 arcmin$^{-2}$ to a \lya\ flux limit of $2 \times 10^{-17}$ ergs
s$^{-1}$ cm$^{-2}$ from a total areal coverage of 700 arcmin$^2$.

Within our adopted cosmological model, an unresolved source at a
redshift $z=8.8$ is 2.7 times fainter than an identical source viewed
$z=5.7$. A further factor of 0.5 must be applied to account for the
different fraction of total flux measured by the 3\arcsec\ diameter
apertures applied by H04 and the 0\farcs7 diameter apertures employed
in the current study. The H04 flux limit therefore corresponds to an
approximate limit of $3.5 \times 10^{-18}$ ergs s$^{-1}$ cm$^{-2}$
when transposed to a redshift $z = 8.8$ and employing 0\farcs7
diameter apertures, i.e., assuming that no additional evolution
occurs, the $z=5.7$ sources observed by H04 are sufficiently bright to
be observed within our NIR selected survey. Applying the surface
density of confirmed $z=5.7$ \lya\ emitting galaxies to the areal
coverage of the current NIR survey, indicates a probability to detect
a \z9 galaxy of 0.12 -- assuming no evolution between the two
redshifts. This `volume shortfall' indicates that the current NIR
selected survey will have to be extended by up to eight times the
currently sampled area in order to realistically probe the no
evolution scenario.

The present narrow $J$--band selected survey has successfully
demonstrated that the sensitivity required to detect putative \z9
\lya\ emitting galaxies emission has been achieved for the case where
no evolution occurs between the redshifts $z=5.7$ and $z=8.8$.
However, it is clear that, in order to place stronger limits on the
space density of such sources, the areal coverage of the current study
must be extended. These observations are currently in progress and
will be presented in a future paper.

\section*{Acknowledgements}

The authors wish to acknowledge the members of the ESO ISAAC
instrument team for their exceptional support in preparing and
executing this programme. The authors further wish to thank J.P. Kneib
and P.C. Hewett for useful comments during the preparation of this
paper.

\appendix

\section[]{The emission line selection function for the NB119 filter and the volume sampled as a function of {\lya} emission luminosity}
\label{app}

The survey sensitivity to \lya\ emission at a redshift \z9\ may be
quantified by considering the volume sampled as a function of \lya\
emission luminosity. In order to constrain the volume averaged \lya\
emission at redshifts \z9, a detailed understanding of the emission
line selection function generated by the NB119 filter is required,
i.e. the probability to identify an emission line of given total flux,
velocity width and central wavelength.

\begin{figure}
\includegraphics[width=84mm,angle=-90.0]{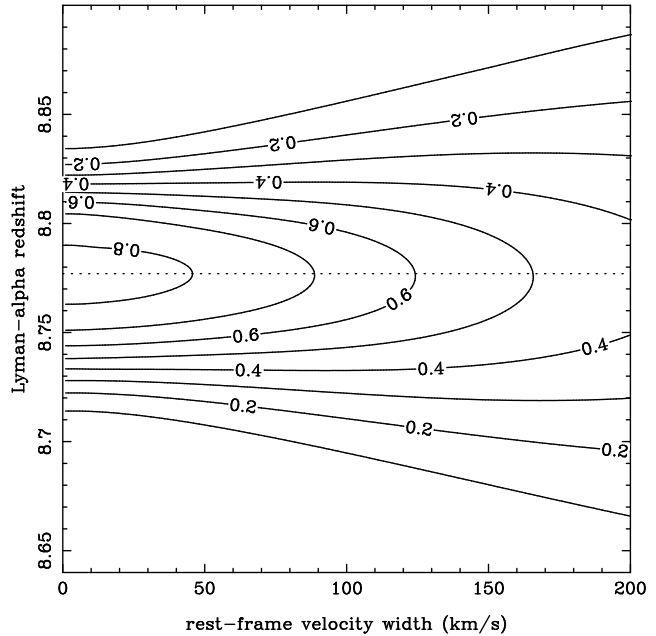}
\caption{Fraction of flux transmitted, $T(z,\sigma_v)$, for a Gaussian
\lya\ emission profile of redshift, $z$, and rest frame velocity
width, $\sigma_v$, convolved with the spectral response of the NB119
filter. The horizontal dotted line indicates the redshift at which
\lya\ emission lies at the central wavelength of the NB119 filter.}
\label{fig_trans_z_sig}
\end{figure}

The flux ratio implied by the $J_s-NB \ge 0.3$ selection criterion
corresponds to an oberved frame \lya\ equivalent width limit of
27{\AA}\footnote{The effective width of the NB119 filter is
85.3{\AA}. We do not consider the attenuation of the putative \z9\
emission spectrum by the intervening IGM in this calculation. However,
as the NB119 spectral response lies very close to the blue wavelength
cut--off of the $J_s$ filter, partial absorption of the $J_s$ flux
compared to the NB119 flux is not considered to be a major systematic
uncertainty.}, or 2.8\AA\ in the rest frame of a $z=8.8$ source. For
comparison, the observed frame EW limit employed by H04 is
approximately 110\AA. The lower limit achieved in the current study
results from a lower narrow-- minus reference--band selection limit
and the use of a narrower filter.  The applied $NB\le25.2$ magnitude
threshold corresponds to a total flux integrated across the NB filter
of $F_{NB} = 3.28 \times 10^{-18} \, \rm ergs \, s^{-1} \,
cm^{-2}$. The above EW threshold therefore indicates that the \lya\
emission flux contributing to this NB flux is approximately one--third
of the total received flux, i.e. $f_{\rm Ly\alpha} = 1.1 \times
10^{-18} \, \rm ergs \, s^{-1} \, cm^{-2}$. The flux sensitivity
calculation is applicable to continuum sources of spectral slope
similar to the mean slope of bright calibration sources within the HDF
South (stars and bright galaxies), i.e. $J_s-NB=0$ (Section
\ref{sec_fluxcal}). However, computation of the emission line
selection function requires the colour term describing the fractional
difference in flux received from an emission line of particular
velocity width and central wavelength (assuming a Gaussian line
profile) compared to the mean calibration source. The NB119 filter
transmits approximately 85{\%} of the total flux received from a
narrow emission line located at the central wavelength of the
filter. The fraction of the incident flux received from a Gaussian
emission line of varying velocity width and central wavelength,
$T(\lambda_c,\sigma_v^\prime)$, may be computed by convolving the
emission line profile with the spectral response of the NB119 filter,
i.e.
\begin{equation}
{
T(\lambda_c , \sigma_v^\prime) = \frac{\int_0^\infty F(\lambda - \lambda_c , \sigma_v^\prime) S(\lambda) \, {\rm d \lambda}} {\int_0^\infty F(\lambda - \lambda_c , \sigma_v^\prime) \, {\rm d \lambda}},
}
\end{equation}
where $F(\lambda - \lambda_c,\sigma_v^\prime)$ is a Gaussian line
profile of central wavelength, $\lambda_c$, and observed--frame
velocity width, $\sigma_v^\prime$. The NB119 filter response is given
by $S(\lambda)$.  The quantities of central wavelength and velocity
width are readily expressed as redshift, $z$, and rest frame velocity
width, $\sigma_v$, assuming that the emission is {\lya}. Contours
describing $T(z,\sigma_v)$ are displayed in Figure
\ref{fig_trans_z_sig}. We do not consider the case where the {\lya}
emission profile is non--Gaussian.

\begin{figure}
\includegraphics[width=84mm]{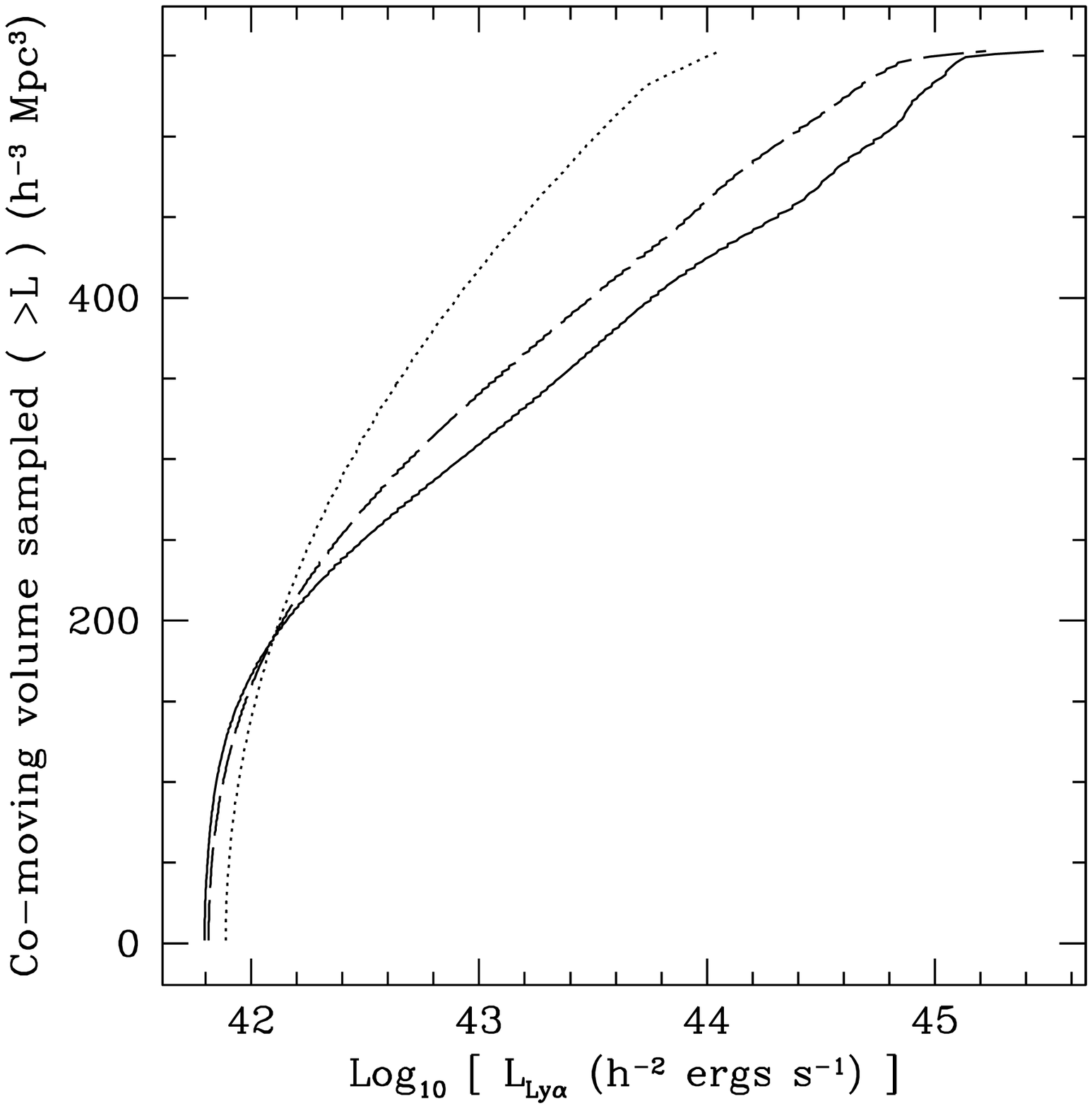}
\caption{Co--moving volume sampled as a function of \lya\ emission
luminosity. Three values of the rest frame \lya\ velocity width are
displayed; $\sigma_v = 20$ (solid line), 50 (dashed line) and 100
(dotted line) kms$^{-1}$.}
\label{fig_lv_samp}
\end{figure}

As noted above, a source satisfying $J-NB \ge 0.3$ and $NB \le 25.2$
corresponds to a transmitted \lya\ emission flux $f_{\rm Ly\alpha} \ge
1.1 \times 10^{-18} \, \rm ergs \, s^{-1} \, cm^{-2}$.  The incident
\lya\ flux, $f_{\rm Ly\alpha}^{\prime}$, is equal to this transmitted
flux divided by the transmitted fraction, i.e.
\begin{equation}
{
f_{\rm Ly\alpha}^{\prime} = f_{\rm Ly\alpha} / T(z,\sigma_v).
}
\end{equation}
The flux limit for a line of fixed rest frame velocity width
corresponds to a varying {\lya} emission luminosity limit as a
function of redshift within the assumed cosmological model according
to ${\rm L_{Ly\alpha}}(z) = 4 \pi \, d_L^{\, 2}(z) f_{\rm
Ly\alpha}^{\prime}(z, \sigma_v)$, where $d_L(z)$ is the luminosity
distance as a function of redshift. The co--moving volume sampled as a
function of {\lya} emission luminosity, $\rm V(L_{Ly\alpha})$ is then
expressed as
\begin{equation}
{\rm V( L_{Ly\alpha})} = b \, {\rm d \Omega} \int_0^{\infty} {\rm H}({\rm
L_{Ly\alpha}}) ({\rm d}V(z)/{\rm d}z) \, {\rm d}z,
\end{equation}
where $b$ corrects for the photometric completeness ($b=0.9$),
d$\Omega$ is the survey solid angle, and
\begin{eqnarray}
{\rm H}({\rm L_{Ly\alpha}}) &=& 1 \; \mbox{where} \; {\rm L_{Ly\alpha}} \ge
{\rm L_{Ly\alpha}}(z, \sigma_v)\\ \nonumber 
&=& 0 \; \mbox{otherwise}.
\end{eqnarray}
Figure \ref{fig_lv_samp} displays the volume sampled as a function of
{\lya} emission luminosity by the NB119 image presented in this
paper. Three rest frame velocity width values are considered to
demonstrate the dependence of the luminosity versus volume sampled
values upon the assumed velocity width. The analysis indicates that
the NB survey selection criteria sample a volume of approximately 340
$h^{-3}$ Mpc$^{3}$ to a {\lya} emission luminosity of $10^{43}$
$h^{-2}$ ergs s$^{-1}$ in the case where the rest frame velocity
width of the source is $\sigma_v = 50 \rm \; kms^{-1}$.

\bsp

\label{lastpage}

\end{document}